\documentclass[journal, 10pt, twocolumn]{IEEEtran}

\usepackage{amsmath,amssymb,amsfonts}
\usepackage{amsthm}
\usepackage{algorithm}
\usepackage{algorithmic}

\usepackage{soul}
\usepackage{subcaption}
\usepackage[pdftex]{graphicx}
\usepackage{xcolor}

\newcommand\unitvec[1]{\ensuremath{\hat{\mathbf #1}}}
\renewcommand\vec{\ensuremath{\mathrm{vec}}}

\newcommand{\A}{\textnormal{A}} 
\newcommand{\B}{\textnormal{B}}  
\newcommand{\bw}{\mathbf{w}}
\newcommand{\bW}{\mathbf{W}}

\newcommand{\bJ}{\mathbf{J}}

\newcommand{\be}{\mathbf{e}}
\newcommand{\ba}{\mathbf{a}}

\newcommand{\bru}{\unitvec{r}}
\newcommand{\bA}{\mathbf{A}}

\usepackage{lipsum}

\theoremstyle{plain}
\newtheorem{proposition}{Proposition}

\begin{document}

\title{Energy-Efficient Design of Broad Beams\\ for Massive MIMO Systems}

\author{Sven~O.~Petersson,
        and~Maksym~A.~Girnyk,~\IEEEmembership{Member,~IEEE}
\thanks{S.~O.~Petersson is with Ericsson Research, Gothenburg, Sweden. M.~A.~Girnyk is with Ericsson Research, Stockholm, Sweden.}
}

\maketitle

\begin{abstract}
Massive MIMO is a promising air interface technique for 5G wireless communications. Such antennas offers capabilities to utilize channel correlations to create beams suitable for user specific transmissions. Most of the prior research is devoted to user-specific transmission in such systems, where a narrow beam is formed towards a user to improve its reception. Meanwhile, public-channel transmission to multiple users at once by means of broad beams has been understudied. In this paper, we introduce the concept of \emph{array-size invariant} (ASI) beamforming, enabling generation of broad beams from very large antenna arrays where all elements are transmitting at maximum power. The ASI technique offers the possibility to achieve a perfectly flat array factor by exploiting the additional degree of freedom coming from polarization diversity. The technique is applicable to uniform linear and rectangular arrays and is characterized by low-complexity beam synthesis. The usefulness of the ASI beamforming is illustrated by means of a numerical example showing the design of a cell-specific broad beam for public-channel transmission in a realistic network deployment.
\end{abstract}



%
\IEEEpeerreviewmaketitle


\section{Introduction}
\label{sec:Introduction}
%
%
%
%

\IEEEPARstart{O}{ne} of the main goals of mobile communication systems have, for many years, been to increase user data rates and system throughput. One of the means for making this happen has been the increase of the number of antennas in order to more efficiently utilize the spatial domain. Various transmission schemes for dedicated channels for exchange of user-specific data have been investigated and standardized in 5G New Radio (NR), see~\cite{asplund2020advanced} for an overview. 

Meanwhile, another important aspect of multi-antenna technology that has received less attention is transmission in the absence of channel state information (CSI), often referred to as \emph{broad-beam} transmission. Such transmission is suitable for public channels, such as, e.g., Physical Downlink Control Channel (PDCCH) and Physical Broadcast Channel (PBCH), for broadcasting cell-specific reference and synchronization signals. Typical requirements on such transmission are a broad radiation pattern and equal power per antenna. Unfortunately, designing beams with such characteristics is not straightforward. It is well known that feeding all elements in an antenna array with the same amplitude and phase results in a narrow beam, potentially much narrower than the angular region in which the served users are located. For instance, 4G Long-Tem Evolution (LTE) specifies \emph{discrete Fourier transform} (DFT) based codebooks~\cite{love2003equal} for user-specific beamforming. 

One way to make a beam broader is to apply amplitude tapering. In the most extreme case, this reduces to transmission from a single active element. However, since practical transmitters employ a single power amplifier (PA) per antenna, this results in a significant reduction in output power due to poor utilization of power resources. Several amplitude-tapering techniques have been proposed in literature. For instance, a beamforming method proposed in~\cite{yang2012random} combines a basis pattern with low gain variations over angle and a phase sequence to ensure equal average power in each direction. A numerical method proposed in~\cite{zhang2017energy} minimizes the gap with the ideal beam pattern, derived via the Parseval identity, using an iterative multi-convex optimization algorithm. An eigenvalue-based optimization approach for codebook beam broadening is proposed in~\cite{raghavan2016beamforming}. Another broad beamforming method, based on minimization of the transmit power, is proposed in~\cite{xiong2017broad}. Furthermore, in~\cite{fan2018flat}, a broad beam is designed by means of Riemannian optimization. Semi-definite programming is applied to broad beam design in~\cite{su2017semidefinite}.

Nevertheless, it is proven in~\cite{qiao2016broadbeam} that the \emph{only} solution able to achieve constant transmitted power for all angles, at a given time instant, is the trivial solution where only one out of all antennas is active. By allowing for a small variation in transmitted power over the angles, the authors of~\cite{qiao2016broadbeam} were able to find beamforming vectors where all antennas radiate. Nevertheless, similarly to the aforementioned algorithms, the obtained solution exhibits mediocre utilization of PAs due to large variations in the amplitude of the elements in the beamforming vector. 

A power-efficient alternative to the above methods is to apply phase tapering. That is, optimize the phases of the excitation weights, while keeping their amplitudes fixed. This, however, typically results in a power pattern with significant ripple. For instance, a method based on Zadoff-Chu sequences, proposed in~\cite{meng2015omnidirectional}, designs beamforming weights that ensure that the signal powers are equal at a set of discrete angles, as many as there are antennas in the array. In~\cite{sayidmarie2013synthesis}, a method for beam broadening was proposed by drawing numerous random weight phases and picking the broadest pattern, satisfying certain conditions. Another broadening approach was proposed in~\cite{intel2016codebook}, where a broadening function is applied to a reference beamforming vector, typically based on a DFT vector. By optimizing parameters of the broadening function, it is possible to increase the angular span of the beam, while reducing the ripple to some acceptable level. Nevertheless, all of the above approaches result in ripple in the shape of the beam.

We note that most of the state-of-the-art references assume that the antenna array is operating with a \emph{single polarization}. However, since the beginning of 4G the majority of antennas, practically used for mobile communications, are \emph{dual-polarized}. That is, an antenna array typically operates with a pair of orthogonal polarizations. Furthermore, most antenna systems are \emph{not} designed with truly omnidirectional coverage in mind. Instead, they are built to cover an angular sector corresponding to a cell.\footnote{For example, in a conventional three-sector deployment, a cell has the angular width of only $120^\circ$.}

Considering the above observations, this paper rests on a slightly different problem formulation, where the total transmitted power, i.e., the sum of powers from two orthogonal polarizations, defines the radiation pattern. This novel approach to beamforming, referred to as \emph{dual-polarization} beamforming, was recently introduced in~\cite{petersson2019power}. Such a formulation allows us to develop a beamforming technique that, as it turns out, makes it possible to find beamformers that result in total power patterns that are constant for all angles at any instant in time. 

The proposed technique is referred to as \emph{array-size invariant} (ASI) beamforming, and it is based on successive \emph{expansion} of an antenna array, while preserving its radiation pattern. The technique ensures excellent power utilization due to constant transmission power across all antennas. Moreover, the technique is applicable to one- and two-dimensional arrays and is characterized with low
complexity (logarithmic in the array size). Finally, it is practically useful since most of the real-world antenna arrays are indeed dual-polarized and could readily adopt the proposed technique. 

The paper's contributions can be summarized as follows:
\begin{itemize}
    \item We introduce the concept of ASI beamforming which allows one to increase the size of an antenna array, while preserving the total radiation pattern of the original array, as well as ensuring that all the available power at all the elements of the array is utilized.
    \item We propose computationally-efficient algorithms for synthesizing perfectly broad beams (without any ripple) for both one- and two-dimensional uniform arrays of arbitrarily large size.
    \item We showcase the usefulness of the proposed ASI beamforming technique by providing a practically-relevant example of designing a broad beam for public-channel transmission in a realistic network deployment.
\end{itemize}

The remainder of the paper is organized as follows. Section~\ref{sec:Antenna arrays} presents system model, as well as considered antenna arrays. Section~\ref{sec:Broad beam generation} discusses the fundamental requirements for the broad beam beamforming concept, as well as presents the main results of the paper: methods for generating broad beams for one- and two-dimensional arrays. Section~\ref{sec:Numerical examples} illustrates the obtained findings with numerical examples. Finally, Section~\ref{sec:Conclusion} presents the conclusions of the study.

The following notations are used throughout the paper. Upper-case and lower-case boldface letters respectively denote matrices and column vectors. Hereafter, $\mathbf[\cdot]^T$ denotes transpose, $\mathbf[\cdot]^*$ denotes conjugate, $[\cdot]^H$ denotes Hermitian transpose, $\mathbf{I}_N$ denotes the $N\times N$ identity matrix, and $\mathbf{J}_N$ denotes the $N\times N$ exchange matrix, i.e., a matrix with ones on the anti-diagonal. The Kronecker product of two matrices  $\mathbf{A}$ and  $\mathbf{B}$ is denoted $\mathbf{A}\otimes\mathbf{B}$. Vectorization of matrix $\mathbf{A}$ is denoted as $\vec(\mathbf{A})$.

\section{System Model}
\label{sec:Antenna arrays}

\begin{figure}[t!]
\centering
\includegraphics[width = \columnwidth]{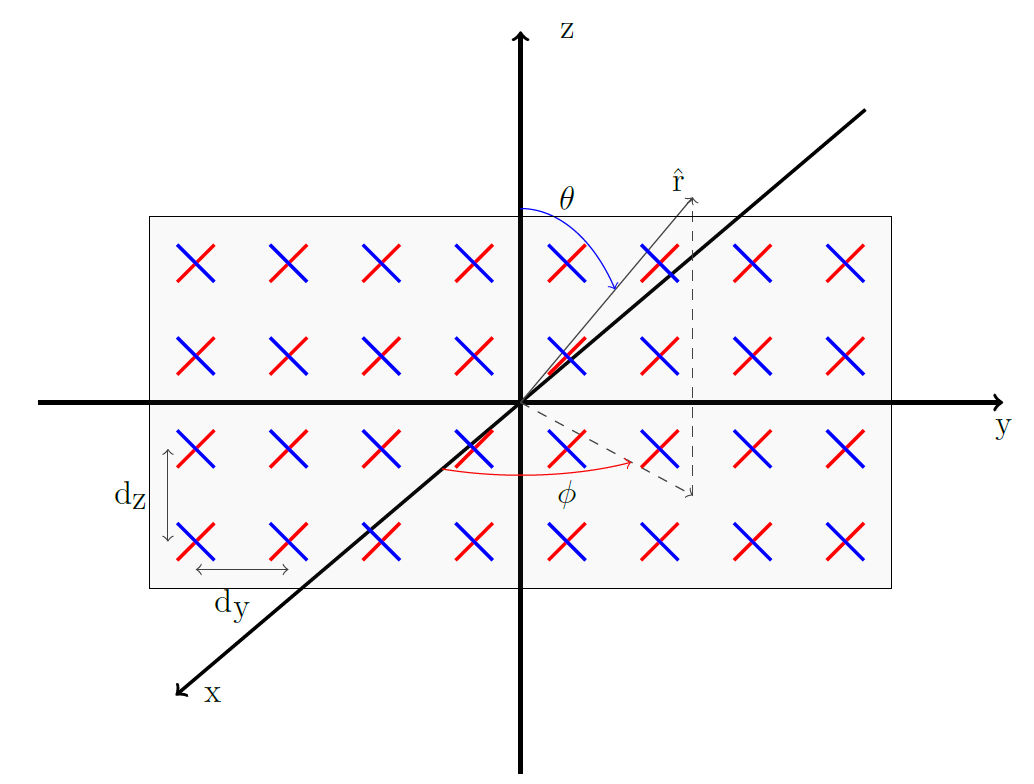}
\caption{Dual polarized array.  $M\times N$-element URA.}
\label{fig:Arrays}
\end{figure}

The antenna arrays considered in this paper are of two types,  uniform linear arrays (ULAs) and uniform rectangular arrays (URAs). It is assumed that all elements in the arrays are identical dual-polarized antenna elements. It is further assumed that the embedded elements are ideal, meaning that the power patterns for the two polarizations are identical and that the element polarizations are perfectly orthogonal. Fig.~\ref{fig:Arrays} shows an illustration of a URA $M\times N$-elements, for which element positions, polarizations and element orientation are indicated. The URA collapses into an $N$-element ULA when we consider the one-dimensional case.

\subsection{Uniform Linear Array}
\label{sec:Antenna arrays ULA}
The steering vector for an $N$-element ULA along the y-axis, in direction \unitvec{r},  is represented by~\cite{balanis2016antenna}

\begin{equation} 
\mathbf{a}(\unitvec{r}) = [ 1, e^{j\psi_\textnormal{y}}, e^{j2\psi_\textnormal{y}}    \cdots, e^{j(N-1)\psi_\textnormal{y}}  ]^T, \label{eq:ULA_steering_vector}
\end{equation}

where \unitvec{r} is a direction vector of unit length, defined as
\begin{equation}
\unitvec{r} =  [\sin\theta \cos\phi,\;\sin\theta\sin\phi,\;\cos\theta]^T, \label{eq:unit_vector} 
\end{equation}
and 
\begin{equation}
\psi_\textnormal{y} = -2\pi d_\textnormal{y}\sin\theta\sin\phi, 
\end{equation}

with $\phi$ being the azimuth angle, $\theta$ being the zenith angle and $d_\textnormal{y}$ being the element spacing in wavelengths along the array axis.\footnote{Note that we here have defined the ULA as a horizontal array to address the issue of forming broad beams in azimuth. This may be useful in the cases where one needs to cover an entire sector with broadcast information. Nevertheless, the need for broad beams also exists in the elevation domain, e.g., to cover a high-rise building.}

Modern communication systems are 
equipped with dual-polarized antennas with almost orthogonal polarizations.
Hence, one can design a desired beam as a total power radiation pattern, i.e., the sum of the partial, per polarization, power patterns by means of the \emph{dual-polarization beamforming} (DPBF) technique~\cite{petersson2019power}. 
Remarkably, it turns out that
with DPBF it is often possible to find unimodular beamforming weights resulting in a desired broad beam shape for the total power 
pattern, therewith, due to constant-modulus beamformers, 
making efficient use of all the available output power. 

Looking at the total power rather than the power for a given polarization---the conventional approach in the beamforming literature---means that the actual polarization state is actually ignored. At first, this might seem counter-intuitive, however, it can be justified by the polarization at the user equipment (UE) anyway being unknown. The latter is due to the impact of the radio channel, the unknown orientation of the UE and the unknown polarization of the UE antenna(s). To counteract performance losses due to all these effects, diversity reception has become a conventional operation at the receive end. Furthermore, if desired, there is always a possibility to create a second beam with identical total power pattern and orthogonal polarization in all directions.

The total complex electric field radiated by the array, when excited by beamforming weight vectors $\mathbf{w_\textnormal{\A}}$, for polarization A, and $\mathbf{w_\textnormal{\B}}$, for polarization B, is given by

\begin{equation}
\mathbf{e}(\unitvec{r})=
\begin{bmatrix}
    \textnormal{e}_{\textnormal{A}}(\unitvec{r})\\
    \textnormal{e}_{\textnormal{B}}(\unitvec{r})
\end{bmatrix} =
\begin{bmatrix}
   \mathbf{w}_{\textnormal{A}}^T\mathbf{a}(\unitvec{r})\\
    \mathbf{w}_{\textnormal{B}}^T\mathbf{a}(\unitvec{r})
\end{bmatrix} E_{\textnormal{el}}(\unitvec{r}),
\label{eq:E_ULA}   
\end{equation}
where $E_{\textnormal{el}}(\unitvec{r})$ is the electric field radiated by a single antenna element, identical for both polarizations.

 The total power radiation pattern is found by superposition of the power for the orthogonal polarizations as
 \begin{align}
{G}(\unitvec{r}) &= {\mathbf{e}}(\unitvec{r})^H \; {\mathbf{e}}(\unitvec{r})\\ 
\label{eq:PowPatt_DP_ULA} 
&=
A(\unitvec{r}) {G_{\textnormal{el}}(\unitvec{r})},
\end{align}
where $G_{\textnormal{el}}(\unitvec{r})=|{E_{\textnormal{el}}(\unitvec{r})}|^2$ is the element power pattern and $A(\unitvec{r})$ is the array factor. As DPBF involves a sum over two orthogonal polarizations, the conventional definition of array factor does not apply. Instead, it is defined as
\begin{equation}
A(\unitvec{r})=\mathbf{a(\unitvec{r})}^H(\mathbf{w_\textnormal{\A}}^*\mathbf{w_\textnormal{\A}}^T+\mathbf{w_\textnormal{\B}}^*\mathbf{w_\textnormal{\B}}^T)\mathbf{a(\unitvec{r})}.  \label{eq:AF_ULA_def} 
\end{equation}

\subsection{Uniform Rectangular Array}
\label{sec:Antenna arrays URA}
Similarly to the case of ULA, the $(m,n)$th entry in the steering matrix $\mathbf{A}(\unitvec{r})$ for a URA reads

\begin{equation}
\label{eq:ULA_steering_vector_element}
{a}_{m,n}(\unitvec{r}) =  e^{j(n\psi_\textnormal{y} + m\psi_\textnormal{z})},
\end{equation}
where
\begin{align}
\psi_\textnormal{y} &= -2\pi d_\textnormal{y}\sin\theta\sin\phi, \\
\psi_\textnormal{z} &= -\pi d_\textnormal{z}\cos\theta.
\end{align}
The total complex far-field pattern is found, in a similar way as for the ULA, when excited by beamforming weight matrices $\mathbf{W_\textnormal{\A}}$, for polarization A, and $\mathbf{W_\textnormal{\B}}$, for polarization B, as
\begin{equation}
\mathbf{e}(\unitvec{r}) =
\begin{bmatrix}
   \vec^T(\mathbf{W}_{\textnormal{A}})\, \vec(\mathbf{A}(\unitvec{r}))\\
   \vec^T(\mathbf{W}_{\textnormal{B}})\, \vec(\mathbf{A}(\unitvec{r}))
\end{bmatrix}E_{\textnormal{el}}(\unitvec{r}).
\label{eq:E_URA}   
\end{equation}

The total power radiation pattern is, as for the ULA, found by superposition of the power for the orthogonal polarizations, as in~\eqref{eq:PowPatt_DP_ULA}. 
The array factor definition for an URA is similar to the one for a ULA, and is given by 

\begin{align}
A(\unitvec{r})= & \vec(\mathbf{A}(\unitvec{r}))^H(
\vec^*(\mathbf{W}_{\textnormal{A}})\vec^T(\mathbf{W}_{\textnormal{A}}) \nonumber \\
& + \vec^*(\mathbf{W}_{\textnormal{B}})\vec^T(\mathbf{W}_{\textnormal{B}})
                        )\vec(\mathbf{A}(\unitvec{r})).
\label{eq:AF_URA_def} 
\end{align}

\section{Broad beam Generation}
\label{sec:Broad beam generation}

In this paper, a broad beam is referred to as a radiation pattern with limited variation in total radiated power in all directions within an angular sector of interest. A beamforming method that produces such a beam shape shall allow for the following:
\begin{itemize}
\item Creating a desired power pattern (for example, but not limited to, the power pattern of a single element);
\item Using the distributed power resource in an efficient way (ideally by having constant modulus beamforming weights).
\end{itemize}

As an extreme case of the above requirements, a broad beam could be defined as the total power radiation pattern in~\eqref{eq:PowPatt_DP_ULA}, being identical to the power radiation pattern for a single element. Equivalently, the array factor must be flat, i.e.,

\begin{equation}
A(\unitvec{r}) = c\unitvec{r},   \label{eq:AF_req_DP} 
\end{equation}
for any direction $\unitvec{r}$, where $c$ is a constant.

\subsection{Existing approaches}

As mentioned in Sec.~\ref{sec:Introduction}, the subject of finding beamformers that fulfill~\eqref{eq:AF_req_DP} has received scholarly attention over the years. Common to all the prior research is an assumption that beamforming is performed using an array where all elements are identical, with respect to both radiation pattern and polarization, i.e., \emph{single-polarization beamforming} (SPBF). Amplitude and phase tapering are widely used approaches for designing a beam with spatially flat array factor for such arrays. 

With amplitude tapering one can optimize amplitudes and phases of the excitation weights to produce a broad beam shape. Unfortunately, in~\cite{qiao2016broadbeam}, it was proven that the only true solution for the problem of designing such a broad beam by means of SPBF is the trivial one where only one element in the excitation vector is active. This solution is, however, not acceptable in the case of active antennas due to the very poor utilization of the available power resources in the antenna array, especially in massive MIMO deployments. 
Another common method for designing a broad beam with improved power utilization is phase tapering. In this case, the corresponding algorithm optimizes only phases of the excitation weights, while keeping the amplitudes constant. In this way, it ensures the all the available power is utilized. There is, however, a limit of the extent to which phase tapering can optimize the beam shape. Typically, it leads to a certain unavoidable amount of ripple in the power pattern. 

As spotted in~\cite{ericsson2017on}, there is a trade-off between power utilization and beam ripple achievable using amplitude and phase tapering. Yet another possible approach is to address this trade-off by means of multi-objective optimization~\cite{deb2014multi}. Such a solution provides a compromise between the two requirements: on constant beam shape and efficient power utilization.

\subsection{Array-Size Invariant Beamforming}
\label{sec:ASI}

Although providing practically acceptable means for wide-sector coverage, the above existing solutions do not enable the generation of broad beams with a flat array factor. In this section, we introduce the concept of \emph{array-size invariant} (ASI) beamforming that provide a means for generating broad beams of controllable width. The concept is based on the previously reported technique of creating beams with identical power patterns and orthogonal polarizations by means of DPBF~\cite{petersson2019power}. The main idea is to utilize the additional degree of freedom that comes from using orthogonal polarizations when designing the antenna excitation weights resulting in a desired beam shape. The excitation weights can often be designed with no amplitude taper leading to full power utilization.

The central technique of the ASI approach is the successive doubling of the size of the array while preserving its power radiation pattern, which we refer to as the \emph{expansion technique}. Consider, for instance, a ULA with $N$ dual-polarized antennas which has been designed so that its power radiation pattern has a desired shape (e.g., as broad as the pattern of a single element); we refer to such an array as a \emph{protoarray}. The task of the expansion technique is to expand the protoarray into an \emph{expanded array} of a larger size, while preserving the radiation pattern of the former. This is done by designing a second array that would be appended to the protoarray without distorting the resulting power pattern; this second array is referred to as the \emph{companion array}. By appending the companion array to the protoarray, we get an expanded array of size $2N$ with the same power radiation pattern as the former. 

\subsubsection{ULA expansion}

Consider a protoarray of size $N$, excited with beamforming weights $\bw_{1,\A}$ and $\bw_{1,\B}$ for the respective polarizations. In order to expand it, while preserving the radiation pattern, the conditions of the following proposition should hold.
\begin{proposition}
	Let $(\bw_{1,\A}, \bw_{1,\B})$ be a pair of per-polarization excitation vectors applied to a protoarray yielding a desired beam shape. Let, furthermore, $(\bw_{2,\A}, \bw_{2,\B})$ be a pair of per-polarization excitation vectors applied to a companion array appended to the protoarray. The radiation pattern of the expanded array with per-polarization  weights given by
	\begin{align}
	\bw_{\A} &= [\bw_{1,\A}^{T}, \bw_{2,\A}^{T}]^{T},\label{eqn:expandedArrayA_1}\\
	\bw_{\B} &= [\bw_{1,\B}^{T}, \bw_{2,\B}^{T}]^{T},\label{eqn:expandedArrayB_1}
	\end{align}
	preserves the pattern of the protoarray if
	\begin{align}
	\bw_{2,\A} &= -\mathbf{J}_N \bw_{1,\B}^{*},\label{eq:w2A}\\
	\bw_{2,\B} &= \mathbf{J}_N \bw_{1,\A}^{*},\label{eq:w2B}
	\end{align}
\end{proposition}

\begin{IEEEproof}
	First, we note that the radiation pattern of the protoarray is given by
	\begin{equation}
	\label{eqn:protoarray_pattern}
	G_{1} (\bru) =\left[ |\bw_{1, \A}^{T} \ba_{1}(\bru)|^2 + |\bw_{1, \B}^{T} \ba_{1}(\bru)|^2\right]  G_\textnormal{el}(\bru)
	\end{equation} 
	
	When appending the companion array to the protoarray, we obtain an expanded array of size $2N$ with excitation weights given by~\eqref{eqn:expandedArrayA_1} and~\eqref{eqn:expandedArrayB_1}, as well as the steering vector given by 
	\begin{align}
	\mathbf{a}(\unitvec{r})&=[{a}_{1}(\unitvec{r}),   \cdots, {a}_{2N}(\unitvec{r}) ]^T.
	\end{align}
	The radiated electric-field vector of the expanded array is given by the superposition of the fields of the two parts
	\begin{align}
	\be(\unitvec{r})
	&=\be_{1}(\unitvec{r}) + \be_{2}(\unitvec{r})\\
	&=\left[\begin{matrix}
	\bw_{1, \A}^{T} \ba_{1}(\unitvec{r}) + \bw_{2, \A}^{T} \ba_{2}(\unitvec{r})\\
	\bw_{1, \B}^{T} \ba_{1}(\unitvec{r}) + \bw_{2, \B}^{T} \ba_{2}(\unitvec{r})
	\end{matrix}\right] E_\textnormal{el}(\bru).
	\end{align}
	The total radiation power pattern of the expanded array is thus given by
	\begin{align}
	G(\bru)
	&=\be^{H}(\bru)\be(\bru)\\
	&=\|\be_{1}(\bru)\|^2 + \|\be_{2}(\bru)\|^2 + 2 \;\textrm{Re}\left\{\be_{1}^{H}(\bru)\be_{2}(\bru)\right\}.
	\end{align}
	If the electric field vectors radiated from the two subarrays are orthogonal, the last term disappears.
	\begin{align}
	\be_{1}^{H}(\bru) \be_{2}(\bru) =  & \left[ \ba_{1}^{H}(\bru) \bw_{1, \A}^* \bw_{2, \A}^{T} \ba_{2}(\bru) \right. \nonumber \\
	& + \left. \ba_{1}^{H}(\bru) \bw_{1, \B}^* \bw_{2, \B}^{T} \ba_{2}(\bru) \right] G_\textnormal{el}(\bru).
	\end{align}
	We notice that the steering vectors of the companion array and the protoarray are related by 
	\begin{equation}
	\ba_{2}(\bru) =  e^{j N  \psi_\textnormal{y}}\; \ba_{1}(\bru),
	\end{equation} 
	where $\psi_\textnormal{y} = -2\pi \;d_\textnormal{y}\sin\theta\sin\phi$. Next, we plug in the weights given in~\eqref{eq:w2A} and~\eqref{eq:w2B} and, utilizing the fact that
	\begin{equation}
	\bJ_N \ba_{2}(\bru) =  e^{j (N-1)  \psi_\textnormal{y}}\;\ba_{2}^*(\bru),
	\end{equation}
	as well as transposing the second term, we get
	\begin{align}
	\be_{1}^{H}(\bru)&\be_{2}(\bru) =   \left[-\ba_{1}^{H}(\bru) \bw_{1, \A}^* \bw_{1, \B}^{H} \ba_{1}^{*}(\bru)  e^{j (2N-1)\psi_\textnormal{y}}\right. \nonumber\\
	& + \left. \left( \ba_{1}^{H}(\bru) \bw_{1, \A}^* \bw_{1, \B}^{H} \ba_{1}^{*}(\bru) e^{j (2N-1) \psi_\textnormal{y}} \right)^T \right]  G_\textnormal{el}(\bru).
	\end{align}
	since both summands are scalars, the transposition operation does not affect them, and hence the two terms inside the braces cancel each other out, yielding $\be_{1}^{H}(\bru) \be_{2}(\bru) = 0$. Thus, the choice of excitation weights for the companion array in~\eqref{eq:w2A} and~\eqref{eq:w2B} leads to electric fields being orthogonal for every observation angle. The total power radiation pattern is therefore given by
	\begin{align}
	G(\bru) &= 2 \left[ |\bw_{1, \A}^{T} \ba_{1}(\bru)|^2 + |\bw_{1, \B}^{T} \ba_{1}(\bru)|^2\right] G_\textnormal{el}(\bru)\\
	&=2 \; G_{1} (\bru),
	\end{align}
	being a scaled version of the pattern of the protoarray~\eqref{eqn:protoarray_pattern}.
	
\end{IEEEproof}

\begin{figure}[t!]
\centering
\includegraphics[width = 8cm]{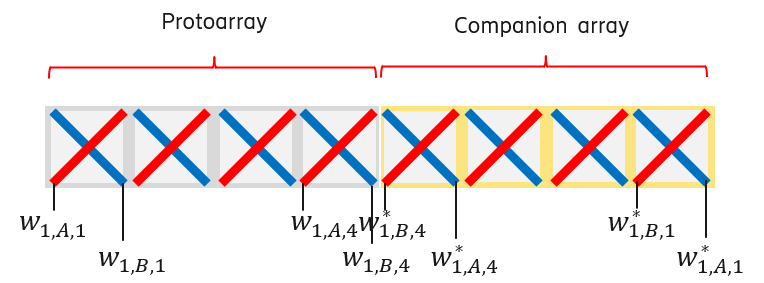}
\caption{Array expansion for ULA by appending a synthesized companion array to the protoarray.}
\label{fig:ASI_1D_single_element}
\end{figure}

The procedure described above is illustrated in Fig.~\ref{fig:ASI_1D_single_element}. It can be conducted successively $k$ times to obtain a $k$-fold array expansion summarized in Algorithm~\ref{alg:expansion1d}. Note that the algorithm has very low computational complexity, $\mathcal{O}(\log N)$.

Remarkably, if the expansion starts from a single antenna element, the entire array will have an array pattern that is a scaled version of the pattern of an individual element. At the same time, all the PAs of the expanded array are fully utilized, and, in principle, there exists no upper limit for the size of such an expanded array.

\begin{algorithm}[t!]
	\caption{ULA Expansion.}
	\label{alg:expansion1d}
	\begin{algorithmic}[1]
		\REQUIRE Size $K = 2^k N$ of the desired large array, $N$-antenna protoarray with a pre-designed pattern given by weights $\bw_{1,A}$ and $\bw_{1,B}$.
		\FOR{$i \gets 1$ to $k$}
		\STATE Compute the excitation weight vectors $\bw_{2,A}$ and $\bw_{2,B}$ of the companion array using~\eqref{eq:w2A} and~\eqref{eq:w2B}.
		\STATE Stack the weight vectors of the companion array and the protoarray as in~\eqref{eqn:expandedArrayA_1} and~\eqref{eqn:expandedArrayB_1} and obtain the excitation weight vectors $\bw_{A}$ and $\bw_{B}$ of an expanded array.
		\STATE $(\bw_{1,A},\bw_{1,B})\leftarrow(\bw_{A},\bw_{B})$, i.e., replace the protoarray with the expanded array.
		\ENDFOR
	\end{algorithmic}
\end{algorithm}

\begin{figure*}[t]
\centering
	\begin{subfigure}{0.45\textwidth}
	    \centering
        \includegraphics[clip, trim = 0.3cm 0.1cm 1.5cm 0cm, height = 5cm]{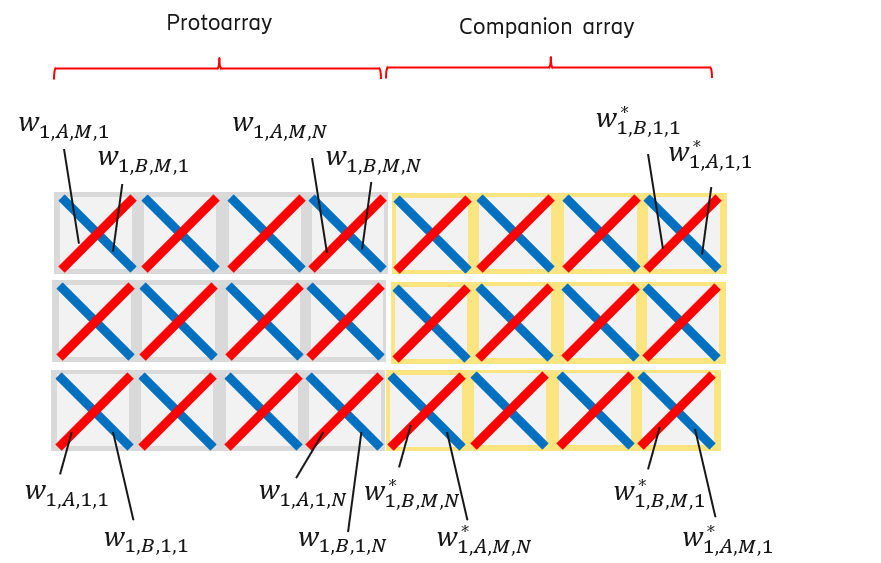}
        \caption{Horizontal expansion.}
        \label{fig:Array_2D_right.png}
	\end{subfigure}%
	\begin{subfigure}{0.55\textwidth}
	    \centering
        \includegraphics[height = 5cm]{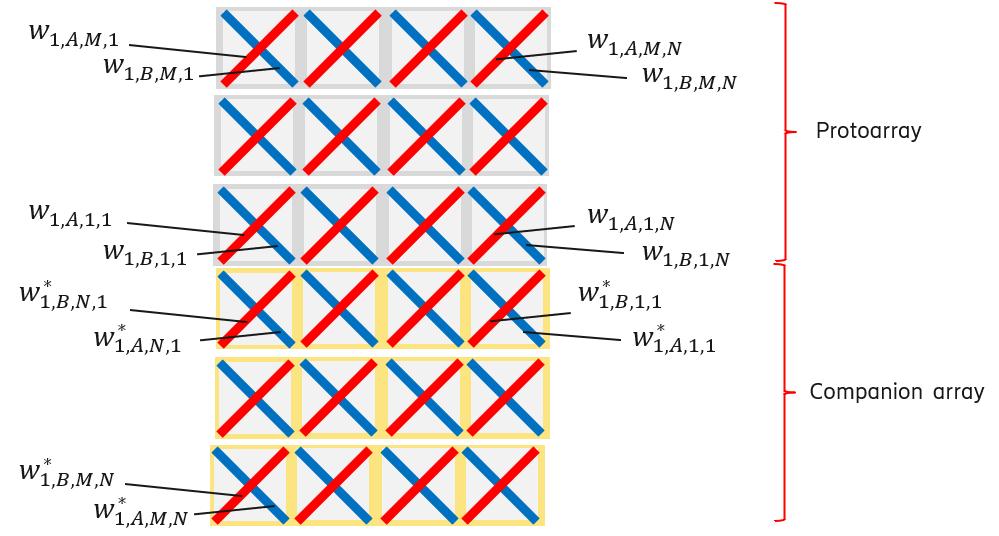}
        \caption{Vertical expansion.}
        \label{fig:Array_2D_vert_below.png}
	\end{subfigure}
\caption{Array expansion for URA by appending the companion array to the protoarray in different ways.}
\label{fig:ASI_2D_single_element}
\end{figure*}

\subsubsection{URA expansion}

Consider a rectangular protoarray of size $N\times M$, excited with beamforming weights $\bW_{1,\A}$ and $\bW_{1,\B}$ for the respective polarizations. In order to expand it with preserving the total power radiation pattern, the conditions of the following should hold.
\begin{proposition}
	Let $(\bW_{1,\A}, \bW_{1,\B})$ be a pair of per-polarization excitation matrices applied to a protoarray yielding a desired beam shape. Let, furthermore, $(\bW_{2,\A}, \bW_{2,\B})$ be a pair of per-polarization excitation vectors applied to a companion array appended to the protoarray. The radiation pattern of the expanded array with per-polarization weights given by one of the ways below
	\begin{align}
	\bW_{\A} &= [\bW_{1,\A}^{T}, \bW_{2,\A}^{T}]^{T}, 
	& \bW_{\B} &= [\bW_{1,\B}^{T}, \bW_{2,\B}^{T}]^{T},\label{eqn:expandedArrayWVer}\\
	\bW_{\A} &= [\bW_{1,\A}, \bW_{2,\A}],
	& \bW_{\B} &= [\bW_{1,\B}, \bW_{2,\B}],\label{eqn:expandedArrayWHor}
	\end{align}
	preserves the pattern of the protoarray if the companion array is excited with
	\begin{align}
		\label{eq:W2A}
		\mathbf{W}_{2,\textnormal{A}}&=-\mathbf{J}_M\mathbf{W}_{1,\textnormal{B}}^*\mathbf{J}_N,\\
		\label{eq:W2B}
		\mathbf{W}_{2,\textnormal{B}}&=\mathbf{J}_M\mathbf{W}_{1,\textnormal{A}}^*\mathbf{J}_N,
	\end{align}
	where $\mathbf{J}_M$ is the exchange matrix of size $M\times M$, $\mathbf{J}_N$ is the exchange matrix of size $N\times N$.
\end{proposition}

\begin{IEEEproof}
	Here we follow the same steps as in the case of ULA. 	The radiated electric-field vector of the expanded array is given by the superposition of the fields of the two parts
	\begin{align}
	\be(\unitvec{r}) = \be_{1}(\unitvec{r}) + \be_{2}(\unitvec{r}),
	\end{align}
	and the total power radiation pattern of the expanded array is thus given by
	\begin{align}
	G(\bru)
	=\|\be_{1}(\bru)\|^2 + \|\be_{2}(\bru)\|^2 + 2 \;\textrm{Re}\left\{\be_{1}^{H}(\bru)\be_{2}(\bru)\right\}.
	\end{align}
	
	The last term disappears when
	\begin{align}
	&\vec^{H}(\bA_{1}(\bru))\vec^*(\bW_{1, \A})  \vec^T(\bW_{2, \A}) \vec(\bA_{2}(\bru))  \nonumber \\
	&+ \vec^{H}(\bA_{1}(\bru))\vec^*(\bW_{1, \B})  \vec^T(\bW_{2, \B}) \vec(\bA_{2}(\bru)) = 0.
	\end{align}
	
	We notice that the steering matrices of the companion array and the protoarray are related as 
	\begin{equation}
	\bA_{2}(\bru) =  \kappa \bA_{1}(\bru),
	\end{equation}
	where $\kappa = \exp(j N \psi_\textnormal{z})$ when the companion array is attached as in~\eqref{eqn:expandedArrayWVer}, and $\kappa = \exp{j M \psi_\textnormal{y}}$ when it is attached as in~\eqref{eqn:expandedArrayWHor}. We plug in the excitation weights given in~\eqref{eq:W2A} and~\eqref{eq:W2B} and, utilizing the fact that 
	\begin{equation}
	\bJ_M \bA_{2}(\bru) \bJ_N = \xi \bA_{2}^*(\bru),
	\end{equation}
	where $\xi =  \exp(j\;[(N-1) \psi_\textnormal{y} + (M-1) \psi_\textnormal{z}])$. Furthermore, transposing the second term, we see that
	\begin{align}
	-&\vec^{H}(\bA_{1}(\bru))\vec^*(\bW_{1, \A})  \vec^H(\bW_{1, \B}) \vec^*(\bA_{1}(\bru)) \kappa \xi \nonumber \\
	+&( \vec^{H}\!(\bA_{1}(\bru))\vec^*(\bW_{1, \A})  \vec^H\!(\bW_{1, \B}) \vec^*(\bA_{1}(\bru))\kappa \xi)^T\!\!\!=\! 0
	\end{align}
	since both summands are scalars. Therefore, $\be_{1}^{H}(\bru) \be_{2}(\bru) = 0$, and the choice of wights for the companion array in~\eqref{eq:W2A} and~\eqref{eq:W2B} leads to electric fields generated by the two arrays being orthogonal for every observation angle. The total radiation power pattern is therefore given by
	\begin{align}
	G (\bru) =&\; 2\left[ |\vec^T(\mathbf{W}_{1, \A}) \vec(\mathbf{A}(\bru)|^2 \right. \nonumber \\ 
 	&+ \left. |\vec^T(\mathbf{W}_{1, \B}) \vec(\mathbf{A}(\bru)|^2 \right]  G_\textnormal{el}(\bru)\\
	=&\;2 \; G_{1} (\bru),
	\end{align}
	being a scaled version of the pattern of the protoarray.
\end{IEEEproof}

The procedure is illustrated in Fig.~\ref{fig:ASI_2D_single_element} on the top of next page. Similarly to the one-dimensional case, the expansion can also be repeated successively $k + l$ times to produce an expanded URA that preserves the radiation pattern of its protoarray, as summarized in Algorithm~\ref{alg:expansion2d}.\footnote{Note that the order of expansion---horizontal, then vertical, or vice versa---does not influence the final radiation pattern, as far as total power pattern is concerned.} Again, the complexity is logarithmic in array sizes, i.e., $\mathcal{O}(\log N + \log M)$.

\begin{algorithm}[t!]
	\caption{URA Expansion.}
	\label{alg:expansion2d}
	\begin{algorithmic}[1]
		\REQUIRE Size $L\times K$ of the desired large array, where $L = 2^{l} M$ and $K = 2^{k} N$. $M\times N$-antenna protoarray with a pre-designed pattern via per-polarization weight matrices $\bW_{1,A}$ and $\bW_{1,B}$.
		
		\texttt{// Horizontal expansion}
		\FOR{$j \gets 1$ to $k$}
    		\STATE Compute the excitation weight matrices $\bW_{2,A}$ and $\bW_{2,B}$ of the companion array using~\eqref{eq:W2A} or~\eqref{eq:W2B}.
    		\STATE Stack the weight matrices of the companion array and the protoarray as in~\eqref{eqn:expandedArrayWVer}, and obtain the excitation weight matrices $\bW_{A}$ and $\bW_{B}$ of an expanded array.
    		\STATE $(\bW_{1,A},\bW_{1,B})\leftarrow(\bW_{A},\bW_{B})$, i.e., replace the protoarray with the expanded array.
		\ENDFOR
		
		\texttt{// Vertical expansion}
		\FOR{$i \gets 1$ to $l$}
		    \STATE Compute the excitation weight matrices $\bW_{2,A}$ and $\bW_{2,B}$ of the companion array using~\eqref{eq:W2A} or~\eqref{eq:W2B}.
    		\STATE Stack the weight matrices of the companion array and the protoarray as in~\eqref{eqn:expandedArrayWVer}, and obtain the excitation weight matrices $\bW_{A}$ and $\bW_{B}$ of an expanded array.
    		\STATE $(\bW_{1,A},\bW_{1,B})\leftarrow(\bW_{A},\bW_{B})$, i.e., replace the protoarray with the expanded array.
		\ENDFOR
	\end{algorithmic}
\end{algorithm}

\section{Numerical Examples}
\label{sec:Numerical examples}

In the previous section, we have shown that by means of the ASI technique it is possible to, for many array sizes, design a beamformer that not only results in a desired total power pattern, but also ensures very efficient utilization of PAs. The companion array, designed according to Algorithms~\ref{alg:expansion1d} and~\ref{alg:expansion2d}, will have the same power utilization as the protoarray. Therefore, the power utilization for the weights, after expansion, will be the same as that for the protoarray. And, given that each antenna element is fed by identical PAs no matter the size of the array, it follows that 

the total output power will increase by a factor of two for each iteration of the algorithm as the array size increases.  

\begin{figure}[!t] 
	\centering
	\begin{subfigure}{0.45\textwidth}
	    \centering
        \includegraphics[width = \textwidth]{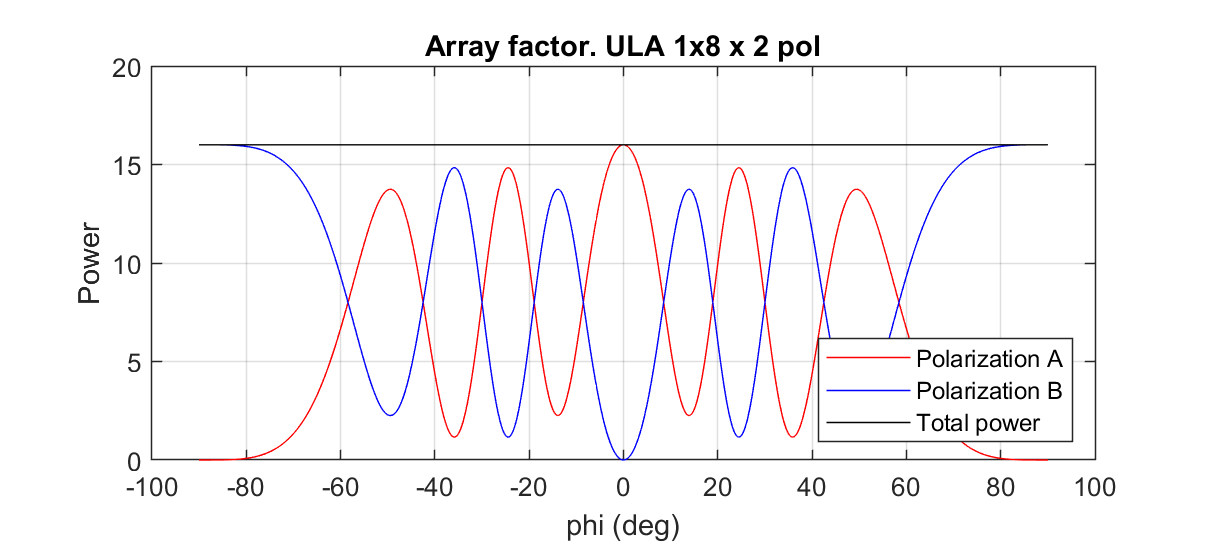}
        \caption{$k=3$.}
        \label{fig:ULA_1x8_arrayfactor.png}
	\end{subfigure}
	\begin{subfigure}{0.45\textwidth}
		\centering
        \includegraphics[width = \textwidth]{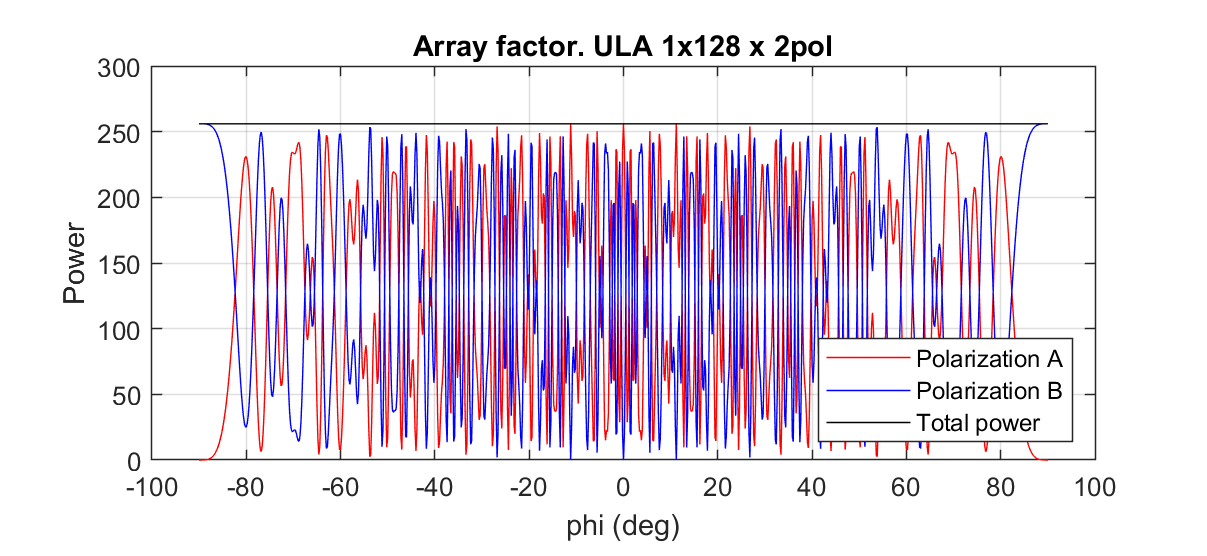}
        \caption{$k=7$.}
        \label{fig:ULA_1x128_arrayfactor.png}
	\end{subfigure}
	\caption{Array factor for single element after $k$-fold one-dimensional expansion.} 
	\label{fig:ULA_arrayfactor}
\end{figure}

Below, we will present several numerical examples illustrating the usefulness of the proposed beamforming method.

\subsection{Spatially flat array factor}
One example which has received a lot of interest in literature in recent years is the design of a constant-modulus beamforming vector whose radiation pattern looks like if it was formed by a single antenna element~\cite{qiao2016broadbeam, su2017semidefinite}. We recall that a beam pattern of a ULA can be written as a product of the array factor and the element pattern. Hence, for the case where the target is a beam pattern similar to that of a single element, we need to find a spatially flat array factor. Interestingly, this array factor can be found for any ULA,
regardless of the element separation. 
Here we demonstrate how beamforming weights can be designed with a simple example.

For that matter, we set the target to be the power pattern formed by a single element. That is, the protoarray is chosen to be a single dual-polarized element with excitation weights given by, e.g.,
\begin{equation}
    \mathbf{w}_{1,\textnormal{A}} = \mathbf{w}_{1,\textnormal{B}} = 1.\label{equ:w_single_element} 
\end{equation}

\subsubsection{Uniform linear array} 
Consider a ULA with $N=8$ dual-polarized antennas. Following the procedure outlined in Algorithm~\ref{alg:expansion1d}, we start with a single dual-polarized element, and expand it $k=3$ times. Thus, we arrive at a pair of excitation vectors of length $2^k=8$, each for its corresponding polarization:
\begin{align}
\bw_{\A} &= [\;1, -1, -1, -1, -1, \hphantom{-} 1, -1, -1]^{T},\label{eqn:expandedArrayA_2}\\
\bw_{\B} &= [\;1, \hphantom{-} 1, -1, \hphantom{-} 1, -1, -1, -1, \hphantom{-} 1]^{T},\label{eqn:expandedArrayB_2}
\end{align}
As can be seen, these excitation vectors are constant modulus so power amplifiers in an active antenna array can be efficiently utilized with no loss due to amplitude taper. 

The array factor corresponding to the excitation weights above is shown in Fig.~\ref{fig:ULA_1x8_arrayfactor.png}. As can be seen from the figure, the patterns in individual polarizations fluctuate with spatial angle. However, in any direction, the power of the two orthogonal polarizations adds up to the same value. Thus, the array factor for the total power is flat. 

A similar example but for a much larger ULA, now with $N=128$ elements is shown in Fig.~\ref{fig:ULA_1x128_arrayfactor.png}. The beamforming weights were found by means of Algorithm~\ref{alg:expansion1d} with $k=7$-fold expansion of a single element. The beamforming vectors are omitted here due to their length. Just as in the previous case, the beamforming vectors are characterized with constant modulus. Therefore, the power loss of more than 8 dB due to amplitude taper reported in~\cite{qiao2016broadbeam} for $N=128$, can  be avoided with the technique presented herein.

\begin{figure*}[!t] 
	\centering
	\begin{subfigure}{0.45\textwidth}
	    \centering
		\includegraphics[width = \textwidth]{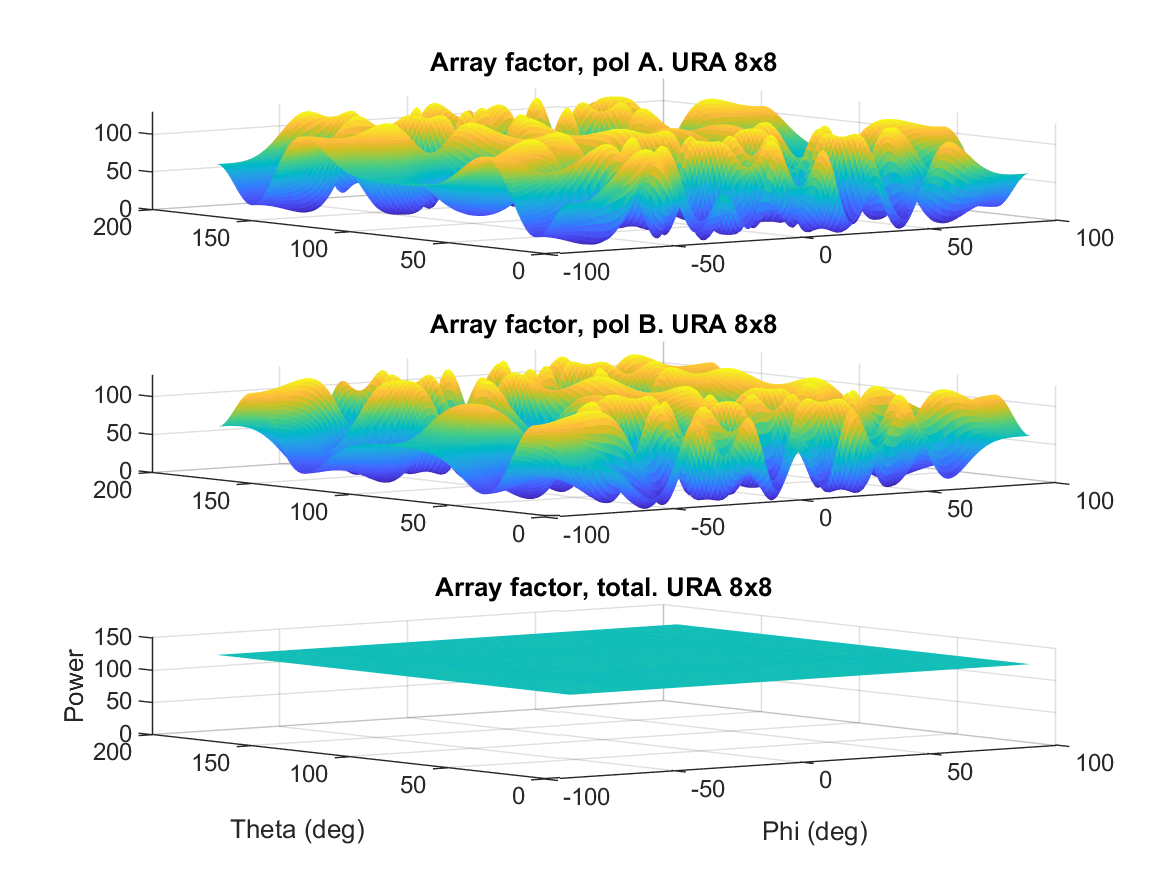}
        \caption{Per-polarization array factors and total array factor.}
        \label{fig:URA_8x8_arrayfactor_cartesian.png}
	\end{subfigure}%
	\begin{subfigure}{0.45\textwidth}
		\centering
		\includegraphics[width = \textwidth]{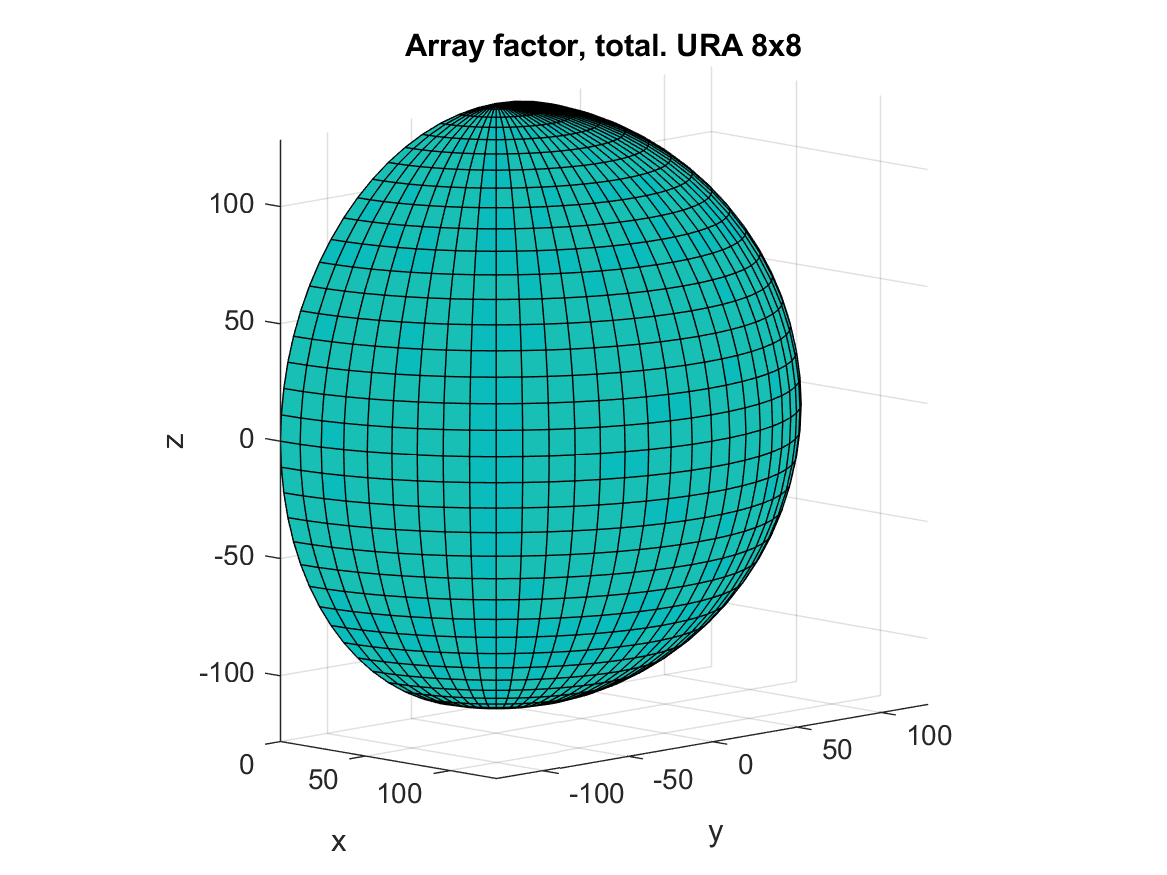}
        \caption{Array factor in polar coordinates.}
		\label{fig:URA_8x8_arrayfactor_polar.png}
	\end{subfigure}
	\caption{Array factor of a single element after two-dimensional expansion by a factor of 8 along each dimension.} 
	\label{fig:URA_8x8_arrayfactor}
\end{figure*}

\subsubsection{Uniform rectangular array} 
Consider a URA with $M\times N = 8\times 8$ antennas. To design the excitation vectors for this case we follow the procedure outlined in Algorithm~\ref{alg:expansion2d}. We start by expanding the single-element protoarray along one dimension. Which dimension to start with does not really matter, but here we have selected to start the expansion along elevation (index $j$) until we have a one-dimensional array of size $8\times1$. The excitation vectors after this step, which are identical to the excitation vectors in~\eqref{eqn:expandedArrayA_2} and~\eqref{eqn:expandedArrayB_2}, now form a new protoarray, which we further expand along the other dimension, i.e., azimuth (index $i$), to form a two-dimensional array. The resulting excitation weight matrices (one per polarization) are shown in~\eqref{eqn:W8x8_polA} and~\eqref{eqn:W8x8_polB}. As can be seen, similarly to the previous case, the excitation matrices are constant-modulus. The array factors per polarization as well as for total power are shown in Fig.~\ref{fig:URA_8x8_arrayfactor_cartesian.png}. The total power array factor (only the front half sphere) is also shown in Fig.~\ref{fig:URA_8x8_arrayfactor_polar.png} in polar coordinates. As can be seen from Fig.~\ref{fig:URA_8x8_arrayfactor}, the array factor for total power is spatially flat and thus the resulting beam pattern will be a scaled version of the element pattern.

\begin{align}
\bW_{\A} = 
\begin{bmatrix} \hphantom{-}1& -1& -1& -1& -1& \hphantom{-}1& -1&  -1\\-1& \hphantom{-}1& \hphantom{-}1& \hphantom{-}1& \hphantom{-}1& -1& \hphantom{-}1& \hphantom{-}1\\-1&  \hphantom{-}1& \hphantom{-}1& \hphantom{-}1& \hphantom{-}1& -1& \hphantom{-}1& \hphantom{-}1\\-1& \hphantom{-}1& \hphantom{-}1& \hphantom{-}1& \hphantom{-}1& -1& \hphantom{-}1& \hphantom{-}1\\-1& -1& \hphantom{-}1& -1& \hphantom{-}1& \hphantom{-}1& \hphantom{-}1& -1\\\hphantom{-}1& \hphantom{-}1& -1& \hphantom{-}1& -1& -1& -1& \hphantom{-}1\\-1& -1&  \hphantom{-}1& -1&  \hphantom{-}1&  \hphantom{-}1&  \hphantom{-}1& -1\\ -1& -1&  \hphantom{-}1& -1&  \hphantom{-}1&  \hphantom{-}1&  \hphantom{-}1& -1\end{bmatrix}\label{eqn:W8x8_polA}\\
\bW_{\B} = 
\begin{bmatrix} \hphantom{-}1& -1& -1& -1& -1& \hphantom{-}1& -1&  -1\\\hphantom{-}1& -1& -1& -1& -1& \hphantom{-}1& -1& -1\\-1&  \hphantom{-}1& \hphantom{-}1& \hphantom{-}1& \hphantom{-}1& -1& \hphantom{-}1& \hphantom{-}1\\\hphantom{-}1& -1& -1& -1& -1& \hphantom{-}1& -1& -1\\-1& -1& \hphantom{-}1& -1& \hphantom{-}1& \hphantom{-}1& \hphantom{-}1& -1\\-1& -1& \hphantom{-}1& -1& \hphantom{-}1& \hphantom{-}1& \hphantom{-}1& -1\\-1& -1& \hphantom{-}1& -1&  \hphantom{-}1& \hphantom{-}1& \hphantom{-}1& -1\\\hphantom{-}1& \hphantom{-}1& -1& \hphantom{-}1& -1& -1& -1& \hphantom{-}1\end{bmatrix}\label{eqn:W8x8_polB}
\end{align}

\subsection{Design of a cell-specific beam}

The above result shows that, even though the conclusions of~\cite{qiao2016broadbeam} hold for SPBF, it is indeed possible achieve omnidirectional transmission by using DPBF (provided that the subelement also has omnidirectional radiation pattern). However, this is rarely required in practice. Instead, one typically needs to design a cell-specific beam with certain characteristics for a specified antenna array. We present an example of such beam design with detailed discussion below.

 \begin{table}
 	\caption{Array parameters considered for the example.}
 	\label{table:Array parameters}
 	\centering
 	\setlength{\tabcolsep}{3pt}
 	\begin{tabular}{p{140pt}p{80pt}}
 		\hline
		Attribute & Value  \\
 		\hline
 		Array size ($M\times N$)& $8\times 8$ elements\\
 		Subarray size & $2 \times1$ elements\\
 		Polarizations & Two orthogonal \\
 		Element pattern (both polarizations) & Gaussian \\
 		Element HPBW, elevation & 90$^\circ$ \\
 		Element HPBW, azimuth & 90$^\circ$ \\
 		Column separation $d_\textnormal{y}$& $0.5 \lambda$ \\
 		Row separation (for URA) $d_\textnormal{z}$ & $0.6\lambda$ \\
 		Target beam pattern shape & Gaussian \\
 		Target HPBW, elevation & 15$^\circ$  \\
 		Target HPBW, azimuth & 65$^\circ$  \\
 		Electrical array tilt & 6$^\circ$  \\
 		Electrical sub-array tilt & 6$^\circ$  \\
 		Inter-site distance & 500 m \\
 		Cell hole radius & 25 m \\
 		Building height & 20 m \\
 		BS height & 25 m \\
 		BS power & 46 dBm\\
 		Number of UE drops per cell & $\sim7400$\\
 		Number of UE antennas & 2, orthogonal polariz.\\
 		UE antenna gain & 0 dBi \\
 		Height outdoor UE & 1.5 m \\
 		Height indoor UE & Uniform within building \\
 		Indoor UEs & 80 \%\ \\
 		Lognormal fading & 6 dB \\
 		Angular spread (elevation and azimuth) & Uniform$(2^\circ,5^\circ)$\\
 		Frequency & 3.5 GHz \\
 		Pathloss model & From TR 38.901~\cite{3gpp2018study} \\
 		Scenario & 3GPP UMa~\cite{3gpp2018study} \\
		\hline
 	\end{tabular}
 	\label{tab1}
 \end{table} 

Consider a typical cellular deployment with three sectors. To cover each sector, an angular range of $120^{\circ}$ should be covered with a single beam. To avoid interference to the neighboring sectors, the beam should have a matching angular width at the level of $-10$ dB with respect to the beam peak. Provided that such a beam has a smooth Gaussian shape, the HPBW of such a beam should be roughly $65^{\circ}$. The pattern of a single cross-pole element typically has HPBW of $90^{\circ}$ or even wider. Therefore, it appears too broad for cell coverage. Meanwhile, the DFT beam produced by a ULA with $N=8$ antennas has HPBW of $\sim 10^{\circ}$, which is too narrow. Therefore, we need to optimize the beam pattern to find an intermediate configuration with the desired HPBW of $65^{\circ}$.

To optimize beamforming weights resulting in a desired beam shape we utilize the robust multi-objective Godlike optimizer~\cite{oldenhuis2010trajectory}. Optimizations here are run with two objectives:
\begin{itemize}
    \item Variance for the difference, in dB, between a target power pattern and the synthesized total power pattern within a certain angular range; 
    \item Power utilization, or equivalently taper loss, defined as
\end{itemize}

\begin{align}
    L_\textnormal{taper}(\mathbf{W}_{\A},\mathbf{W}_{\B}) =&   \frac{2MN\max(|[\vec^{T}(\mathbf{W}_{\A}), \vec^{T}(\mathbf{W}_{\B})]^{T}|^2)}{\|[\vec^{T}(\mathbf{W}_{\A}), \vec^{T}(\mathbf{W}_{\B})]^{T}\|^2}.
    \label{eqn:TaperLoss}   
\end{align}
To find the beamforming matrices for a URA with $M\times N$ dual-polarized elements one could run the optimization for the entire 2D-array. However, that is a difficult and time consuming task due to the large number of parameters. To simplify the former, we note that the total radiation pattern is a combination of the elevation pattern and the azimuth pattern. Therefore, we can optimize beamforming vectors for elevation and azimuth separately, thereby significantly reducing the computational complexity.

The array configuration and deployment of interest are summarized in Table~\ref{table:Array parameters}. Note that in contrast to the azimuth dimension, where UEs are spread widely over the entire $120^{\circ}$-wide sector, in the elevation dimension, UEs are located within a rather narrow angular sector. Hence, a narrow radiation pattern is required in elevation, pointing at the center of the UE distribution, as seen from the BS. In practice, this is often achieved by means of \emph{virtualization}, i.e., partitioning the entire array into subarrays and treating those as array elements, as well as applying a tilt angle. This means that our $M\times N$ array is reduced to an array of size $M/2\times N$ whose elements will be subarrays of size $2\times 1$ with pattern down-tilted by $6^{\circ}$. Again, according to the pattern multiplication property, the total radiation pattern of the array is given by the product of the array factor and the pattern of the subarray.

\begin{figure*}[!t] 
	\centering
	\begin{subfigure}{0.45\textwidth}
	    \centering
		\includegraphics[width = \textwidth]{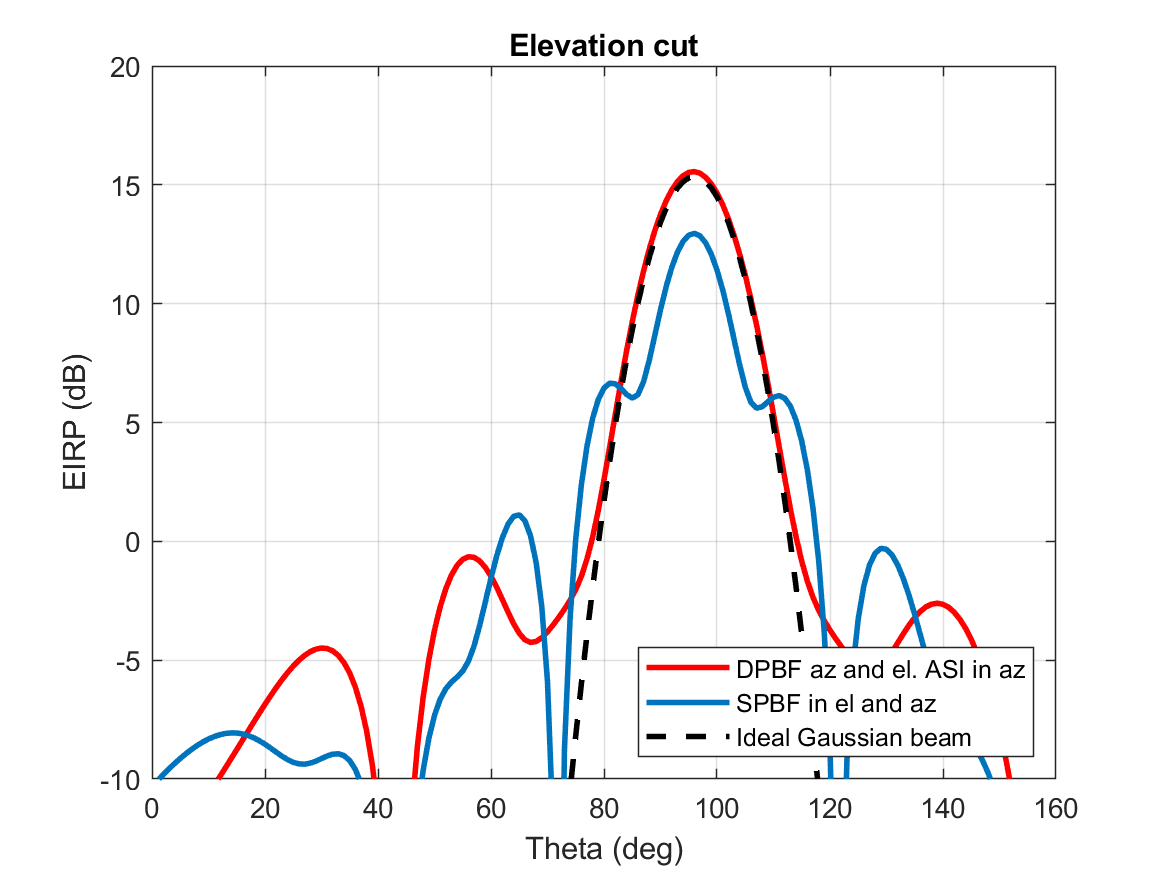}
        \caption{Elevation cut.}
        \label{fig:Elcut.png}
	\end{subfigure}%
	\begin{subfigure}{0.45\textwidth}
		\centering
		\includegraphics[width = \textwidth]{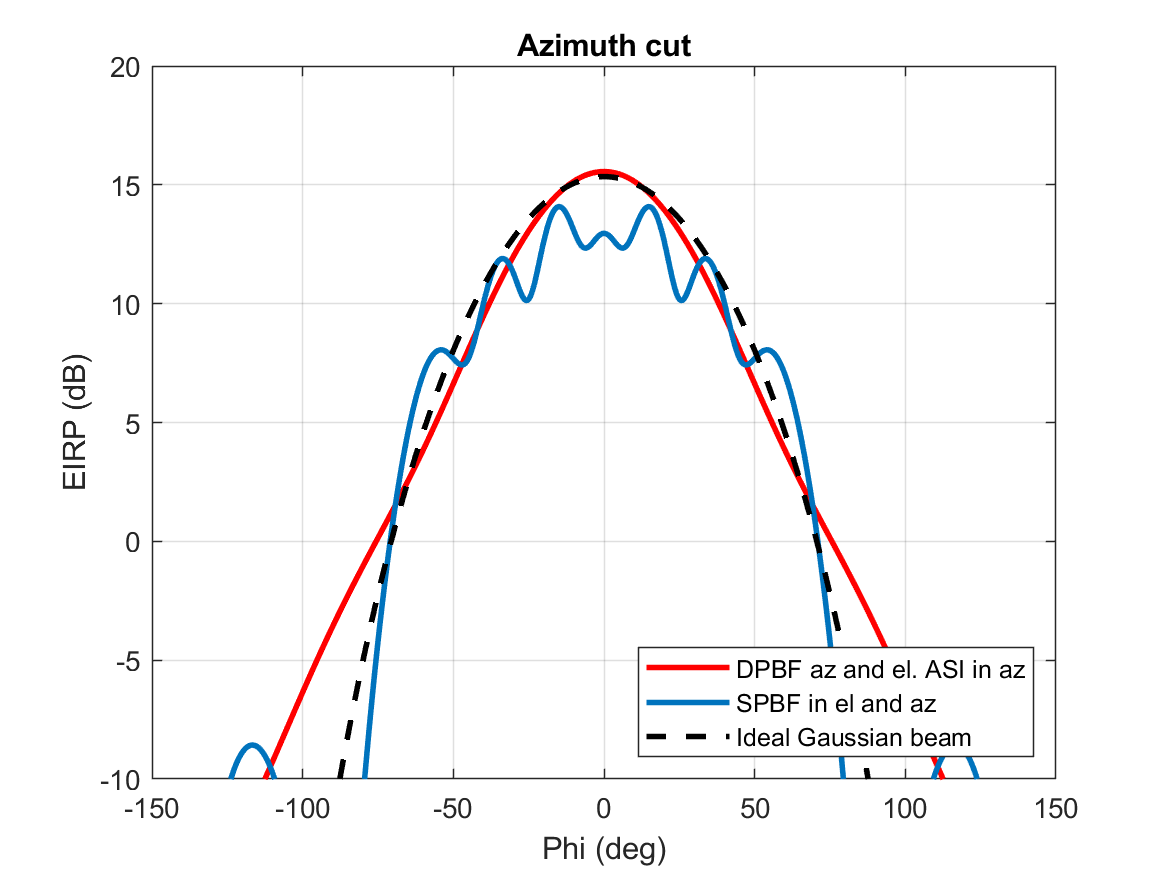}
        \caption{Azimuth cut.}
		\label{fig:Azcut.png}
	\end{subfigure}
	\caption{Elevation and azimuth cuts for optimized beam patterns.} 
	\label{fig:BeamCuts.png}
\end{figure*}

\subsubsection{SPBF radiation patterns} 

For the case of SPBF, the beam pattern for the sector-wide public-channel transmission is optimized separately for elevation and azimuth patterns. The patterns are then combined into a three-dimensional pattern as follows. Assume that $\bw_{\mathrm{y}} \in \mathbb{C}^N$ and $\bw_{\mathrm{z}} \in \mathbb{C}^M$ are optimized beamforming vectors in azimuth and elevation dimensions, respectively. The optimized SPBF beamforming matrix of the entire two-dimensional array is then given by

\begin{equation}
    \bW = \bw_{\mathrm{z}} \otimes \bw_{\mathrm{y}}^T \in \mathbb{C}^{M\times N}.
    \label{eqn:2D_SPBF}
\end{equation}

In elevation, the following SPBF beamforming vector (electrical tilt not included) is used to feed the 4 subarrays,
\begin{equation}
\bw_{\mathrm{z},1} = \begin{bmatrix}\;0.297 - 0.146j\\ 0.271 + 0.256j\\ 0.267 + 0.260j\\ 0.301 - 0.146j\end{bmatrix}.\label{eqn:wz_SPBF}\\
\end{equation}

The corresponding vector feeding the elements is found via
\begin{equation}
\bw_{\mathrm{z}} = \bw_{\mathrm{z},1} \otimes [ 1, 1]^T \in \mathbb{C}^{M}.
\label{eqn:wz_SPBF_elementspace}\\
\end{equation}

In azimuth, the following SPBF beamforming vector is used to feed the 8 columns
\begin{equation}
\bw_{\mathrm{y}} = \begin{bmatrix}\;\hphantom{-}0.189 + 0.189j\\ \hphantom{-}0.093 - 0.165j\\  -0.244 + 0.112j\\ -0.129 + 0.232j\\ -0.129 + 0.232j\\ -0.244 + 0.112j\\ \hphantom{-}0.093 - 0.165j\\  \hphantom{-}0.189 +0.189j\end{bmatrix}.\label{eqn:wy_SPBF}\\
\end{equation}

As both $\bw_{z}$ and $\bw_{z}$ are SPBF vectors the excitation matrix for the two-dimensional array $\bW$,~\eqref{eqn:2D_SPBF}, will also be SPBF. Thus, only half of the elements, and hence PAs, in the array will be in use. For that reason we feed also the orthogonal polarization B using the same excitation matrix as for the first polarization A. Finally, we apply a scaling factor, $\alpha$, to the total excitation matrix, which takes base station (BS) output power from Table~\ref{tab1} and taper loss~\eqref{eqn:TaperLoss} into account,
\begin{equation}
    \alpha = (P_\textnormal{BS}/L_\textnormal{taper})^2/{\|[\vec^{T}(\mathbf{W}_{\A}), \vec^{T}(\mathbf{W}_{\B})]^{T}\|},
\label{eqn:W_scaling_1}
\end{equation}
This scaling factor ensures the correct output power without overloading any power amplifier. In addition, we also need to apply the tilt according to Table~\ref{tab1}.

\subsubsection{DPBF radiation patterns} 
The process for deriving excitation matrices for DPBF is similar to the one for SPBF patterns, but there are some important differences. One is that we now have two matrices, one per element polarization, for both elevation and azimuth beamforming. Another one is that, when creating the matrices for the entire two-dimensional array, one must avoid introducing amplitude variations. 

Here, the elevation pattern is optimized using the Godlike optimizer~\cite{oldenhuis2010trajectory} with the constraints that the excitation weights shall be constant modulus (i.e., phase-only taper) and that the elevation power pattern shall be symmetrical with respect to its peak.

The following DPBF beamforming vector is used to feed the 4 subarrays for polarization A,
\begin{equation}
    \bw_{\mathrm{z,1,A}} = \begin{bmatrix}\;\hphantom{-}0.271 + 0.227j\\ \hphantom{-}0.181 + 0.304j\\-0.091 + 0.342j\\-0.199 + 0.292j\end{bmatrix}.\label{eqn:wzA_DPBF}\\
\end{equation}
The corresponding vector in element space is found via
\begin{equation}
    \bw_{\mathrm{z,A}} = \bw_{\mathrm{z,1,A}} \otimes [ 1, 1]^T \in \mathbb{C}^{M}.
\label{eqn:wzA_DPBF_elementspace}\\
\end{equation}
For polarization B, the beamforming vector is defined as
\begin{equation}
    \bw_{\mathrm{z,B}} = \bw_{\mathrm{z,A}}^*\label{eqn:wzB_DPBF}
\end{equation}
to meet the symmetry requirement for the total power pattern.

In azimuth, a constant-modulus beamforming vector for a protoarray consisting of two dual polarized elements is found via a bisection search, similar to what is done in~\cite{girnyk2020simple}. Requiring that the total power pattern shall be symmetrical, the resulting vectors are
\begin{align}
\bw_{\mathrm{y,A}} &= \begin{bmatrix}\;0.458 - 0.200j\\  0.458 + 0.200j\end{bmatrix},\label{eqn:wyAproto_DPBF}\\
\bw_{\mathrm{y,B}} &= \bw_{\mathrm{y,A}}^*.\label{eqn:wyBproto_DPBF}
\end{align}

Next, these beamforming vectors are expanded according to Algorithm~\ref{alg:expansion1d} to length 4, i.e., $k = 1$, and zero padded to the full length of 8. Avoiding extending the weight to the full length, zero padding instead, is important to avoid introducing amplitude variations when the excitation matrix for the entire two-dimensional array is created (see~\cite{petersson2019power} for more details).

The matrices, one per element polarization, for the entire two-dimensional array are found as
\begin{align}
    \bW_{\mathrm{A}} &= \bw_{\mathrm{z,A}} \otimes \bw_{\mathrm{y,A}}^T-\bw_{\mathrm{z,B}} \otimes \bJ\bw_{\mathrm{y,B}}^H \in \mathbb{C}^{M\times N},\\
    \bW_{\mathrm{B}} &= \bw_{\mathrm{z,A}} \otimes \bw_{\mathrm{y,B}}^T+\bw_{\mathrm{z,B}} \otimes \bJ\bw_{\mathrm{y,A}}^H \in \mathbb{C}^{M\times N},
\end{align}
The weights are furthermore properly normalized with a scaling factor,~\eqref{eqn:W_scaling_1},
to have the correct output power without overloading any PA. In addition, a tilt is applied according to Table~\ref{tab1}.

The radiation patterns for the optimized weights are shown in Fig.~\ref{fig:BeamCuts.png}, alongside the target patterns. Note that 
the elevation cuts in Fig.~\ref{fig:Elcut.png} are shown after applying the electrical tilt according to Table~\ref{tab1}. Also note that the patterns are plotted in terms of equivalent isotropically radiated power (EIRP). For these plots the beamforming weights are normalized, such that the output power is 1~W minus taper loss.

\begin{figure}[t!]
    \centering
    \includegraphics[width = 8cm]{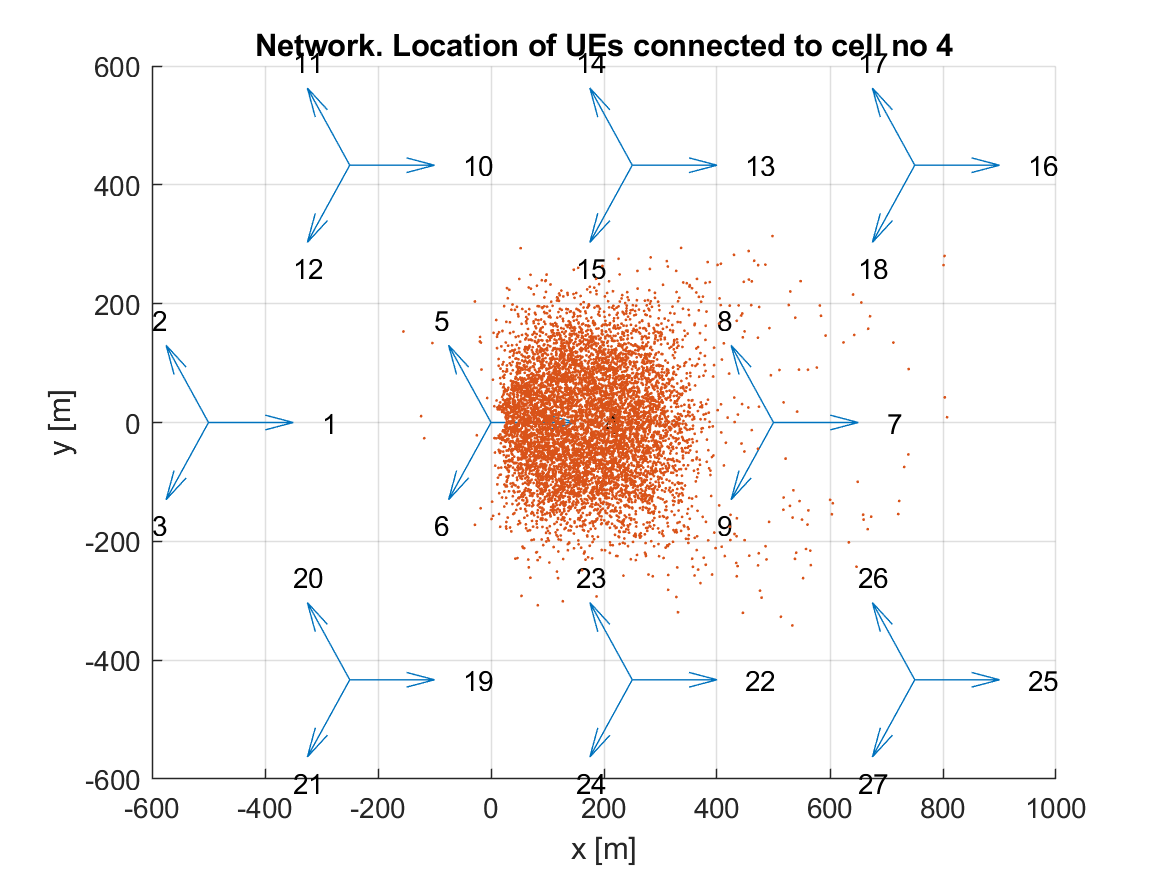}
    \caption{Network deployment and UE locations for DPBF.}
    \label{fig:Sample_UEloc_Jul22}
\end{figure}

\subsubsection{Received signal power}

To illustrate the performance gain obtained from the proposed technique, we simulate a system with a total of 27 cells in an Urban Macro (UMa) environment~\cite{3gpp2018study}. All BSs are equipped with the same antenna array and hence have the same beam pattern. Users are dropped with uniform distribution in the horizontal plane, although avoiding the cell hole of 25 m around the BS. 80\% of the users are located indoors and for these the height is given by a uniform distribution within [0, building height]. Users are served by the cell with the lowest pathloss. An example of network layout is shown in Fig.~\ref{fig:Sample_UEloc_Jul22}. The figure shows the positions of 9 three-sector sites. The blue arrows in the figure indicate the boresight directions of the antenna arrays, while the orange dots indicate the UE locations for users connected to a particular cell.

\begin{figure}[t!]
\centering
\includegraphics[width = 8cm]{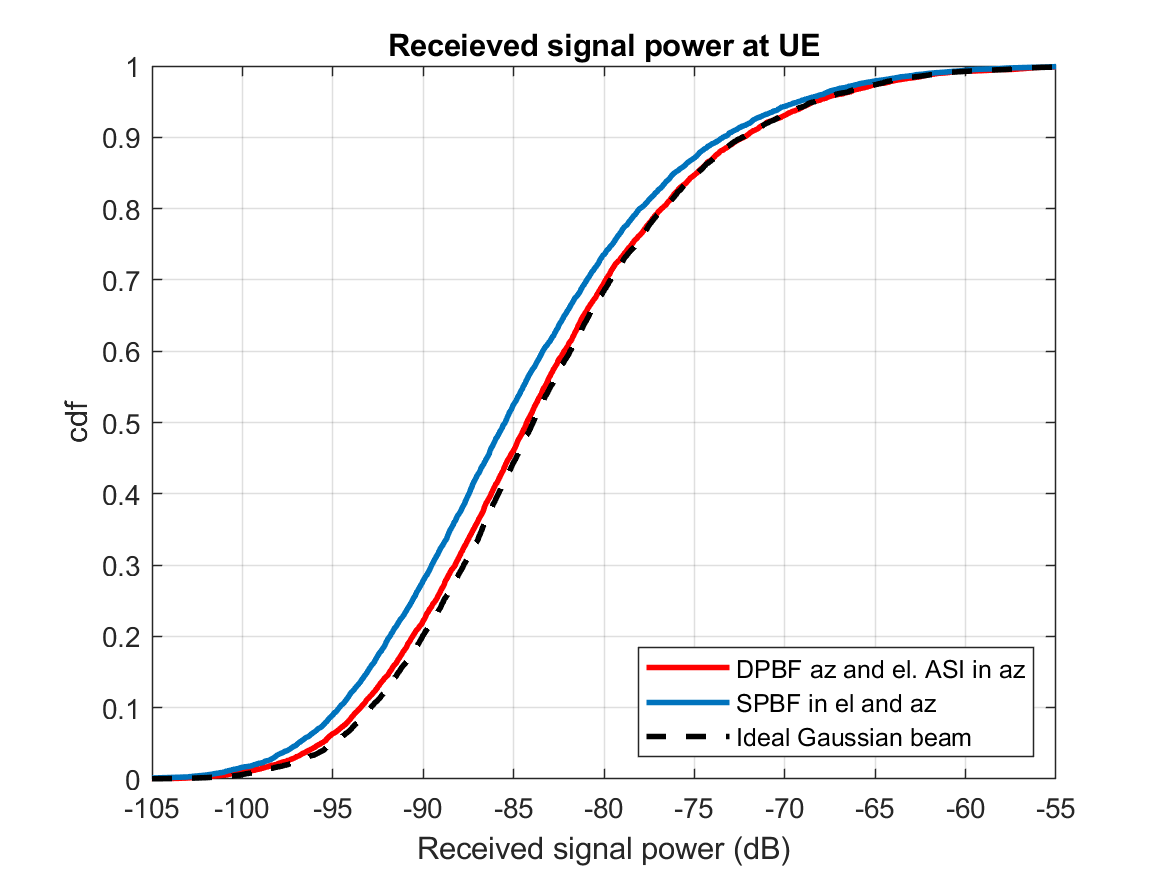}
\caption{Performance comparison between SPBF and DPBF.}
\label{fig:URA_8x8_ServingSignal}
\end{figure}

Received signal power, that is the desired signal total power received by the UE, is used as a performance metric whose cumulative distributions for the different beamforming methods are shown in Fig.~\ref{fig:URA_8x8_ServingSignal}. The figure shows that the proposed DPBF-based ASI beamformer outperforms the SPBF solution and approaches the upper bound, the ideal Gaussian beam. The former shows a gain of 1-1.5 dB in received signal power. This is partly explained by the taper loss of 1.1 dB for the SPBF solution, as was observed in~\eqref{eqn:TaperLoss}. Furthermore, the gain comes also partly due to the shape of the resulting power pattern. All in all, the figure demonstrates the superiority of the proposed DPBF- and ASI-based beam design.

\section{Conclusion}
\label{sec:Conclusion}

In this paper, we have proposed the technique of array-size invariant (ASI) beamforming that allows us to expand an antenna array, while preserving its total radiation pattern. The ASI technique is based on successive doubling of a dual-polarized array, exploiting the additional degree of freedom coming from the two available polarizations. It is based on phase-only beamforming, and hence guarantees full power utilization due to constant-modulus weights. The technique is applicable to uniform linear and rectangular arrays and is characterized with low complexity for beam synthesis. We have propose two computationally-efficient algorithms for designing broad beams for uniform arrays of arbitrarily large size. Our simulations the benefits of the proposed ASI beamforming technique vs. single-polarization beamforming for designing a cell-specific broad beam for public-channel transmission in a realistic network deployment.

\section*{Acknowledgment}

The authors would like to thank their colleagues at Ericsson Research for stimulating discussions throughout the course of this work.

\ifCLASSOPTIONcaptionsoff
  \newpage
\fi



%

\bibliographystyle{IEEEtran}
\bibliography{refs}




%







\end{document}